\documentclass[
    ,final            
    ,sort&compress    
  ]
  {aipproc}

\layoutstyle{6x9}

\begin{document}
\def\gsim{\mathrel{\rlap{\lower2pt\hbox{\hskip0pt$\sim$}}\raise3pt\hbox{$>$}}}
\vspace{-1cm}
\title{pp Elastic Scattering at LHC and Nucleon Structure (Conference Report)\footnote{Presented by M. M. Islam at the Conf. on Intersections of Particle and Nuclear Physics
(N.Y., May 2003), the 8$^{th}$ Int. Wigner Symposium (CUNY, N.Y., May 2003)
and the 10$^{th}$ Blois Workshop (Helsinki, June 2003).
Details of the paper have appeared in: Mod. Phys. Lett. 18(2003)743; hep-ph/0210437.}}
\author{M. M. Islam}{
  address={Department of Physics, University of Connecticut, USA}
}

\author{R. J. Luddy}{
  address={Department of Physics, University of Connecticut, USA}
}

\author{A. V. Prokudin}{
  address={Department of Theoretical Physics, University of Torino and INFN, Torino, Italy}
}

\begin{abstract}
High energy elastic $pp$ differential cross section at LHC at the
c.m. energy 14 TeV is predicted using the asymptotic behavior
of $\sigma$$_{tot}$(s) and $\rho$(s), and the measured $\bar{p}p$
differential cross section at $\sqrt{s}$ =546 GeV. The
phenomenological investigation has progressively led to an effective
field theory model that describes the nucleon as a chiral bag
embedded in a quark-antiquark condensed ground state.
The measurement of $pp$ elastic scattering at LHC up to
large |t| $\gsim$ 10 GeV$^{2}$ by the TOTEM group will be crucial
to test this structure of the nucleon.
\end{abstract}

\maketitle


\vspace{-.4cm}

High energy $pp$ and $\bar{p}p$  elastic scattering have been measured at
the CERN ISR[1]\nocite{Nagy:1979} and SPS Collider[2-3] 
\nocite{Bozzo:1984}\nocite{Bernard:1986} over a wide range of energy and momentum 
transfer: $\sqrt{s} $ = 23-630 GeV and |t| = 0-10 GeV$^2$.
These measurements have been followed by Fermilab
Tevatron measurement[4-5]\nocite{Amos:1990}\nocite{Abe:1994} 
of $\bar{p}p$  at $\sqrt{s}$ = 1.8 TeV and |t| = 0-0.5 GeV$^2$.
Such a large experimental effort naturally presents us with
the following questions:
\flushleft
\vspace{-.2cm}
1. What do we learn from these experiments about NN interactions
at high energies?\\

2. What insight do we get from them about the physical structure of the nucleon?

\hspace{.3cm}  These questions have now assumed greater significance because of the Large
Hadron Collider (LHC) currently being built at CERN.  One of the first
experiments planned at LHC called TOTEM (Total and Elastic Measurement)
will measure $pp$ elastic $d\sigma \over dt$ in the near forward
direction at an unprecedented c.m. energy $\sqrt{s}$ = 14 TeV. \\

\hspace{.3cm}  My collaborators and I have been studying high energy $pp$, $\bar{p}p$ elastic
scattering for some time[6-8]\nocite{Heines:1981}\nocite{Islam:1984}\nocite{Islam:1987}.
Our initial phenomenological investigation led us to the following description.
The nucleon has an outer cloud and an inner core (Fig. 1).  High energy
elastic scattering is primarily due to two processes (Fig. 2): 1) a glancing
collision where the outer cloud of one nucleon interacts with that of the
other giving rise to diffraction scattering; 2) a hard (or large |t|)
collision where one nucleon core scatters off the other core via vector
meson $\omega$ exchange, while their outer clouds overlap and interact independently.
In the small |t| region diffraction dominates, but the hard scattering
takes over as |t| increases.

\hspace{.3cm}  Let me present an example from our recent calculations.  The solid curve in
Fig. 3 is our calculated $d\sigma \over dt$ for $\bar{p}p$ scattering at
$\sqrt{s}$ =546 GeV.  The dotted curve is the differential cross section
due to diffraction alone, while the dot-dashed curve is that due to the hard
scattering alone.  As we can see, diffraction dominates in the small |t|
region, but falls off rapidly as |t| increases, and the hard scattering
takes over.  The interference between the diffraction and the hard
scattering produces the dip.  The experimental data are from SPS
Collider[2]\nocite{Bozzo:1984}.  The thick dashed curve in Fig. 3
is our calculated $pp$ elastic $d\sigma \over dt$ at $\sqrt{s}$ = 500 GeV,
which is currently being measured at RHIC in the small |t| region[9]\nocite{Guryn:2001}.\\

\hspace{.3cm}  We describe diffraction scattering using the impact parameter representation:

\vspace{-.1cm}
\flushright
$T_D (s,t)=i\,p\,W\int_0^\infty {b\;db\;J_0 } (b\,q)\Gamma_{D}(s,b)$;\hspace{4cm} (1)
\flushleft
$q$ is the momentum transfer ($q = \sqrt{|t|}$) and $\Gamma_{D}(s,b)$ is 
the profile function, which is related to the eikonal function $\chi_{D}(s,b)$:
$\Gamma_{D}(s,b) = 1 - exp(i \chi_{D}(s,b))$. We take $\Gamma_{D}(s,b)$ to
be an even Fermi profile function:

\vspace{-.2cm}
\flushright
$\Gamma_{D}(s,b) = g(s)$ [$ 1 \over 1 + exp((b-R)/a)$ + $ 1 \over 1 + exp(-(b+R)/a)$ -1 ].\hspace{2 cm}(2)
\flushleft
\vspace{-.2cm}

The parameters $R$ and $a$ are energy dependent: $R$ = $R_0 + R_1(ln s -$ $i\pi\over 2$ $)$,
$a$ = $a_0 + a_1(ln s -$ $i\pi\over 2$ $)$; $g(s)$ is a complex crossing even
energy-dependent coupling strength.\\

\hspace{.3cm}  Our hard scattering amplitude is of the form

\vspace{-.2cm}
\flushright
$T_H (s,t)\;\sim \;\exp [i\;\chi _D (s,0)]\;s\;\frac{F^2(t)}{m_\omega ^2 -t}$.\hspace{4 cm}(3)
\flushleft

The t-dependence is the product of two form factors and
the $\omega$ propagator.  It shows that $\omega $ probes two density
distributions corresponding to the two form factors. The density
distributions represent the nucleon cores.  The factor of s originates
from spin 1 of $\omega$.  The factor $e^{i\;\chi_D (s,0)}$  represents
absorptive correction due to diffraction scattering.
The diffraction amplitude obtained by us satisfies a number of general properties
associated with the phenomenon of diffraction:\\

\vspace{.1cm}
1.  $\sigma_{tot}(s) \sim (a_{0} + a_{1} ln s)^2$ (Froissart-Martin bound)\\
2.  $\rho (s) \simeq \frac{\pi a_1}{a_0 + a_1 ln s}$ (derivative dispersion relation)\\
3.  $T_{D}(s,t) \sim i\; s\; ln^{2}s\; f(|t| ln^{2}s)$  (AKM scaling)\\
4.  $T_{D}^{\bar{p}p}(s,t) = T_{D}^{pp}(s,t)$      (crossing even)\\

\hspace{.3cm}  Our present approach is different from our earlier one[8]\nocite{Islam:1987},
where we fitted known $d\sigma \over dt$ at different energies using complex
energy-dependent parameters.  Our goal now is to obtain the asymptotic behavior
and the approach to the asymptotic behavior of the
elastic scattering amplitude, so that we can predict the $pp$
differential cross section at $\sqrt{s}$ = 14 TeV.  To this end, we
require the energy-dependent parameters to describe quantitatively
the asymptotic behavior and the approach to the asymptotic behavior
of total cross section $\sigma_{tot}(s)$ and
$\rho(s) = \frac{Re T(s,0)}{Im T(s,0)}$ as known from dispersion
relation calculations.  Furthermore, we require them to describe well the
measured $\bar{p}p$ elastic differential cross section at 546 GeV[2]\nocite{Bozzo:1984}.
Here are the results of our calculations
of $\sigma_{tot}(s)$ (Fig. 4), $\rho(s)$ (Fig. 5), and $d\sigma \over dt$
at $\sqrt{s}$ = 546 GeV (Fig. 3) shown earlier.  We find a satisfactory
description.  Once the parameters are determined, we can test our model by
predicting $d\sigma \over dt$ at higher energies where experimental data are
available.  Fig. 6 shows our prediction at $\sqrt{s}$ = 1.8 TeV for $\bar{p}p$
compared with the Tevatron data[4-5]\nocite{Amos:1990}\nocite{Abe:1994}. Fig. 7 shows
our prediction for $\bar{p}p$ elastic scattering at $\sqrt{s}$ = 630 GeV,
where large |t| data are available from SPS Collider[3]\nocite{Bernard:1986}.
These tests indicate that the model provides a reasonably quantitative
description of high energy elastic scattering.\\

\hspace{.3cm}  We now proceed to predict $pp$ elastic $d\sigma \over dt$ at LHC at the
c.m. energy 14 TeV (Fig. 8).  The solid curve is our predicted differential cross
section.  The dashed curve represents the prediction by the impact-picture
model of Bourrely et al. and the dot-dashed curve represents that by the Regge
pole-cut model of Desgrolard et al.[10-13]\nocite{Bourrely:1985}\nocite{Bourrely:1996}\nocite{Desgrolard:1994}
\nocite{Buenerd:1996}.  The latter models predict typical diffraction
oscillations in the large |t| region, while our model predicts smooth
fall-off of $d\sigma \over dt$ for |t| > 1.5 GeV$^{2}$.  The dotted line
in Fig. 8 represents schematically the expected change in our model
in the behavior of $d\sigma \over dt$ from Orear fall-off: $d\sigma \over dt$
$\sim$ $e^{-a\sqrt{|t|}}$ to a power fall-off: $d\sigma \over dt$
$\sim$ $t^{-10}$ due to quark-quark scattering.

\hspace{.3cm}  Our phenomenological investigation progressively led us to an effective field theory model
that describes the nucleon structure.  This development began with a
criticism of our model which was the following:  The hard scattering
amplitude in our model (Eq.(3)) has a factor of $s$ from spin 1
of $\omega$, and the $s$ and $t$
dependence of this amplitude shows that $\omega$ behaves as an elementary
vector meson.  On the other hand, at such high energies one would expect
$\omega$ to Reggeize and $s$ be replaced by $s^{\alpha_{\omega}(t)}$, where
$\alpha_{\omega}(t)$ is the $\omega$ trajectory. $\alpha_{\omega}(t)$ is
considerably less than 1 at large |t| and therefore this amplitude
should give negligible contribution contrary to our calculations.
However, we noticed that in the non-linear $\sigma$-model
of the nucleon, $\omega$ couples to the baryonic current like
a gauge boson: $g\omega_\mu\hspace{-2pt}J_{B}^{\mu}$, and the baryonic current
is topological:

\vspace{-.2cm}
\flushright
$J_{B}^{\mu} = \frac{\epsilon^{\mu\nu\rho\sigma}}{24 \pi^{2}}
\;\;tr[U^{\dagger}\;\partial_{\nu}U\;U^{\dagger}\;\partial_{\rho}U\; U^{\dagger}\;\partial_{\sigma}U]$ \hspace{4 cm}(4)
\flushleft

What this model says is that it is an effective field theory
model.  But, as long as it holds, baryonic current continues to behave
as a topological current and $\omega$ coupled to it as a gauge
boson continues to behave as a gauge boson, i.e. as an elementary
vector meson.  And we seem to be seeing this behavior.

\hspace{.3cm}  Fortunately, there was a way of testing this conclusion.  From
our $\omega$$NN$ form factor $F(t)$, we can obtain by Fourier transform
the baryonic charge distribution and then derive the pion field
that gives rise to this baryonic charge distribution.  We can compare this
pion field with the pion field obtained in the n.l. $\sigma$-model,
which describes the nucleon as a topological soliton or Skyrmion.  Here is the
result of our analysis: Fig. 9.  The solid curve is our calculated pion field
configuration, or pion profile function $\theta (r)$, while the dotted
and the dashed curves are the pion profile functions from the
n.l. $\sigma$-model.  The curves are consistent with each other, if
we bear in mind that our curve is coming from c.m. energy region $\gsim$ 23 GeV,
while the other curves are coming from an energy region of order 1 GeV.
Furthermore, the r.m.s. radius for the baryonic charge distribution
obtained by us is 0.44 F, while that from the n.l. $\sigma$-model is about 0.5 F.\\

\hspace{.3cm}  We faced another problem at this point.  Even though the
n.l. $\sigma$-model when gauged describes the low energy properties
of the nucleon quite well, it typically predicts a soliton mass
$m_{sol}$ $\sim$ 1500 MeV compared to the nucleon mass $m_{N}$ = 939 MeV
(see, for example,[14]\nocite{Zhang:1994}). We obviously had to confront this
problem of large soliton mass as we were claiming evidence in favor of
the soliton model. To this end, we examined a model
more general than the n.l. $\sigma$-model. The model turns out to be
the linear $\sigma$-model of Gell-Mann and Levy, which is described by
the Lagrangian:

\vspace{.2cm}
$\mathcal{L}=\bar\psi i\gamma^{\mu}\partial_{\mu}\psi +
\frac{1}{2}(\partial_{\mu}\sigma\partial^{\mu}\sigma+
\partial_{\mu}\vec{\pi}\partial^{\mu}\vec{\pi}) -
g\bar{\psi}[\sigma+i\vec{\tau}\cdot\vec{\pi}\gamma^{5}]\psi-
\lambda(\sigma^{2}+\vec{\pi}^{2}-f_{\pi}^{2})^{2}$. (5)
\vspace{.2cm}


The model has $SU(2)_{L}\;\; \times \;\; SU(2)_{R} \;\;\times \;\;U(1)_{V}$ global symmetry and
spontaneous breakdown of chiral symmetry. $\psi$ is the quark field,
$\sigma$ is an isospin-zero scalar field, and $\vec{\pi}$ is an
isovector pseudoscalar field.  The model can be expressed in terms of
right and left quark fields $\psi_{R,L} = \frac{1}{2}(1\pm \gamma^{5})\psi$
by introducing a scalar field $\zeta$ and a unitary field $U$ in the
following way: $\sigma + i  \vec{\tau}\cdot \vec{\pi}=\zeta U, 
\zeta = \sqrt{\sigma^{2}+\vec{\pi}^{2}}, 
U = e^{\frac{i\vec{\tau}\cdot\vec{\phi}}{f_{\pi}}}.$\\

The field $\vec{\phi}$ is the massless Goldstone pion field; $U$ is the
Skyrmion field that gives rise to the topological baryonic current.  In
terms of these fields, Eq.(5) takes the form

\vspace{-.3cm}

\[
\mathcal{L} = \bar\psi_{R}\;i\;\gamma^{\mu}\;\partial_{\mu}\psi_{R} +
    \bar\psi_{L}\;i\;\gamma^{\mu}\;\partial_{\mu}\psi_{L} +
    \frac{1}{2}\partial_{\mu}\zeta\partial^{\mu}\zeta     +
    \frac{1}{4}\zeta^{2}tr[\partial_{\mu}U\;\partial^{\mu}U^{\dagger}]\hspace{3 cm}
\]

\flushright
\vspace{-.4cm}

 $-g \; \zeta \; (\bar\psi_{L}\;U\;\psi_{R} +\bar\psi_{R}\;U^{\dagger}\;\psi_{L}) -
    \lambda \; (\zeta^{2} - f_{\pi}^{2})^{2}$.\hspace{5 cm}(6)
\vspace{-.2cm}
\flushleft

\hspace{.3cm}  In the conventional n.l. $\sigma$-model, one replaces from the very
beginning the scalar field $\zeta$ by its vacuum value $f_\pi$.
Furthermore, one introduces a Weiss-Zumino-Witten anomalous
action term[15]\nocite{Bhaduri:1988}, which arises from the underlying
quark structure of the model.  It is this action that contains
the term $g\omega_{\mu}J_{B}^{\mu}$, which couples $\omega$ to the
topological baryonic current.  The n.l. $\sigma$-model also assumes
that all the important low-energy interactions are in the meson sector.
The only important interaction coming from the quark sector is that
given by the WZW action and no further interaction in the quark
sector needs to be included.  This, of course, leads to a Skyrmion
lying in a non-interacting Dirac sea (Fig. 10).  On the other
hand, we notice from the linear $\sigma$-model that even though
replacing $\zeta$ by its vacuum value $f_{\pi}$ may be reasonable
in the meson sector, completely neglecting it in the quark sector
is questionable, because the $\zeta$ field provides an interaction
between left and right quarks (Eq.(6)). The latter makes the quarks massive
and leads to the spontaneous breakdown of chiral symmetry.
This, of course, means
that we have a soliton lying in an interacting Dirac sea (Fig. 10).
What one finds is that if the scalar field has a critical behavior,
and by this I mean a scalar field that is zero for small distances,
but rises sharply at some distance $R$ to its vacuum value $f_{\pi}$ (Fig. 11),
then the energy of the interacting Dirac sea together with that of the
scalar field is considerably less than that of the non-interacting
Dirac sea[16]\nocite{Islam:1992}.  The system therefore makes a transition
to this lower ground state and significantly reduces its total energy
or mass.  This condensation phenomenon solves the
soliton mass problem and is analogous to superconductivity.
Instead of spin up and down electrons, we have left and right quarks
forming a $q\bar{q}$ condensate.

\hspace{.3cm}  The behavior of the $\zeta$ field shown in Fig. 11 has significant
implications.  First, for $r < R$, $\zeta = 0$ --- quarks are massless,
and we are in a perturbative regime.  Therefore, we end up with the
nonperturbative structure of the nucleon shown in Fig. 12, which shows
that the nucleon is a chiral bag[15,17]\nocite{Bhaduri:1988}\nocite{Hosaka:2001}
embedded in a $q\bar{q}$ condensed ground state.  Second, for momentum
transfer $Q = \sqrt{|t|}$ sufficiently large, one nucleon probes the
other nucleon at an impact parameter $b \; \approx \frac{1}{Q} \; < \; R$,
and therefore in the perturbative regime where $pp$ elastic scattering
originates from valence quark-quark scattering.  The latter has been
investigated by Sotiropoulos and Sterman[18]\nocite{Sotiropoulos:1994} who
concluded that at very large $|t|$, $\frac{d\sigma}{dt} \; \sim \; t^{-10}$
(same as power counting rules).  From our point of view, this means that
as momentum transfer $Q$ increases, there will be a critical value
$Q_{0} \; \approx \; \frac{1}{R}$ beyond which $\frac{d\sigma}{dt}$ will
tend to a power fall-off.  Schematically, the dotted line in Fig. 8
represents this transition from the nonperturbative regime to the
perturbative regime and, in fact, will be a signature of the chiral
phase transition.
\newpage
\begin{center}
\textbf{Concluding remarks}
\end{center}
\vspace{-.2cm}

1.  Our phenomenological investigation has led us to physical aspects of the nucleon which have been proposed and studied by other authors in different contexts.
\vspace{.1cm}

2.  We find that the nucleon is a chiral bag embedded in a quark-antiquark ground state, and this ground state is analogous to a superconducting ground state.  We also find that this structure is described by an effective field theory model --- a gauged Gell-Mann-Levy linear $\sigma$-model. 
\vspace{.1cm}

3.  The experimental study of pp elastic scattering at LHC at $\sqrt{s}$ = 14 TeV by the TOTEM group up to large $|t|$ will be crucial to test this structure of the nucleon.

\vspace{-.6cm}

\bibliographystyle{aipproc}   
\bibliography{islam}

\hyphenation{Post-Script Sprin-ger}
\begin{thebibliography}{18}
\expandafter\ifx\csname natexlab\endcsname\relax\def\natexlab#1{#1}\fi
\providecommand{\enquote}[1]{``#1''}
\expandafter\ifx\csname url\endcsname\relax
  \def\url#1{\texttt{#1}}\fi
\expandafter\ifx\csname urlprefix\endcsname\relax\def\urlprefix{URL }\fi

\bibitem[Nagy et~al.(1979)]{Nagy:1979}
Nagy, E., et~al., \emph{Nucl. Phys.}, \textbf{B150}, 221 (1979).

\bibitem[Bozzo et~al.(1984)]{Bozzo:1984}
Bozzo, M., et~al., \emph{Phys. Lett.}, \textbf{B147}, 385 (1984),
  \hspace{-5pt}; \textbf{B155}, 197 (1985).

\bibitem[Bernard et~al.(1986)]{Bernard:1986}
Bernard, D., et~al., \emph{Phys. Lett.}, \textbf{B171}, 142 (1986).

\bibitem[Amos et~al.(1990)]{Amos:1990}
Amos, N., et~al., \emph{Phys. Lett.}, \textbf{B247}, 127 (1990).

\bibitem[Abe et~al.(1994)]{Abe:1994}
Abe, F., et~al., \emph{Phys. Rev.}, \textbf{D50}, 5518 (1994).

\bibitem[Heines and Islam(1981)]{Heines:1981}
Heines, G.~W., and Islam, M.~M., \emph{Nuovo Cimento}, \textbf{61A}, 149
  (1981).

\bibitem[Islam et~al.(1984)]{Islam:1984}
Islam, M.~M., Fearnley, T., and Guillaud, J.~P., \emph{Nuovo Cimento},
  \textbf{81A}, 737 (1984).

\bibitem[Islam et~al.(1987)]{Islam:1987}
Islam, M.~M., Innocente, V., Fearnley, T., and Sanguinetti, G., \emph{Europhys.
  Lett.}, \textbf{4}, 189 (1987).

\bibitem[Guryn(2001)]{Guryn:2001}
Guryn, W., \emph{Nucl. Phys. (Proc. Suppl.)}, \textbf{B99}, 299 (2001).

\bibitem[Bourrely et~al.(1985)]{Bourrely:1985}
Bourrely, C., Soffer, J., and Wu, T.~T., \emph{Phys. Rev. Lett.}, \textbf{54},
  757 (1985).

\bibitem[Bourrely et~al.(1996)]{Bourrely:1996}
Bourrely, C., et~al., \enquote{Quarter Century of Rising Total Cross Sections,}
  in \emph{Frontiers in Strong Interactions}, edited by P.~Chiappetta,
  M.~Haguenauer, and J.~T.~T. Van, Editions Frontieres, 1996, p.15.

\bibitem[Desgrolard et~al.(1994)]{Desgrolard:1994}
Desgrolard, P., Giffon, M., and Predazzi, E., \emph{Z. Phys.}, \textbf{C63},
  241 (1994).

\bibitem[Buenerd(1996)]{Buenerd:1996}
Buenerd, M., \enquote{The TOTEM project at LHC,} in \emph{Frontiers in Strong
  Interactions}, edited by P.~Chiappetta, M.~Haguenauer, and J.~T.~T. Van,
  Editions Frontieres, 1996, p.\hspace{1pt}437.

\bibitem[Zhang and Mukhopadhyay(1994)]{Zhang:1994}
Zhang, L., and Mukhopadhyay, N.~C., \emph{Phys. Rev.}, \textbf{D50}, 4668
  (1994).

\bibitem[Bhaduri(1988)]{Bhaduri:1988}
Bhaduri, R.~K., \emph{Models of the Nucleon: from Quarks to Soliton},
  {Ad\-di\-son-Wes\-ley}, Reading, Massachusetts, 1988.

\bibitem[Islam(1992)]{Islam:1992}
Islam, M.~M., \emph{Z. Phys.}, \textbf{C53}, 253 (1992).

\bibitem[Hosaka and Toki(2001)]{Hosaka:2001}
Hosaka, A., and Toki, H., \emph{Quarks, Baryons, and Chiral Symmetry}, World
  Scientific, 2001.

\bibitem[Sotiropoulos and Sterman(1994)]{Sotiropoulos:1994}
Sotiropoulos, M.~G., and Sterman, G., \emph{Nucl. Phys.}, \textbf{B425}, 489
  (1994).

\end{thebibliography}

\vspace{.2cm}

\textbf{Figures 1 - 12:}  Explanations of the figures are given in the main body of the paper.

\begin{figure}[b]
  \includegraphics[height=.24\textheight]{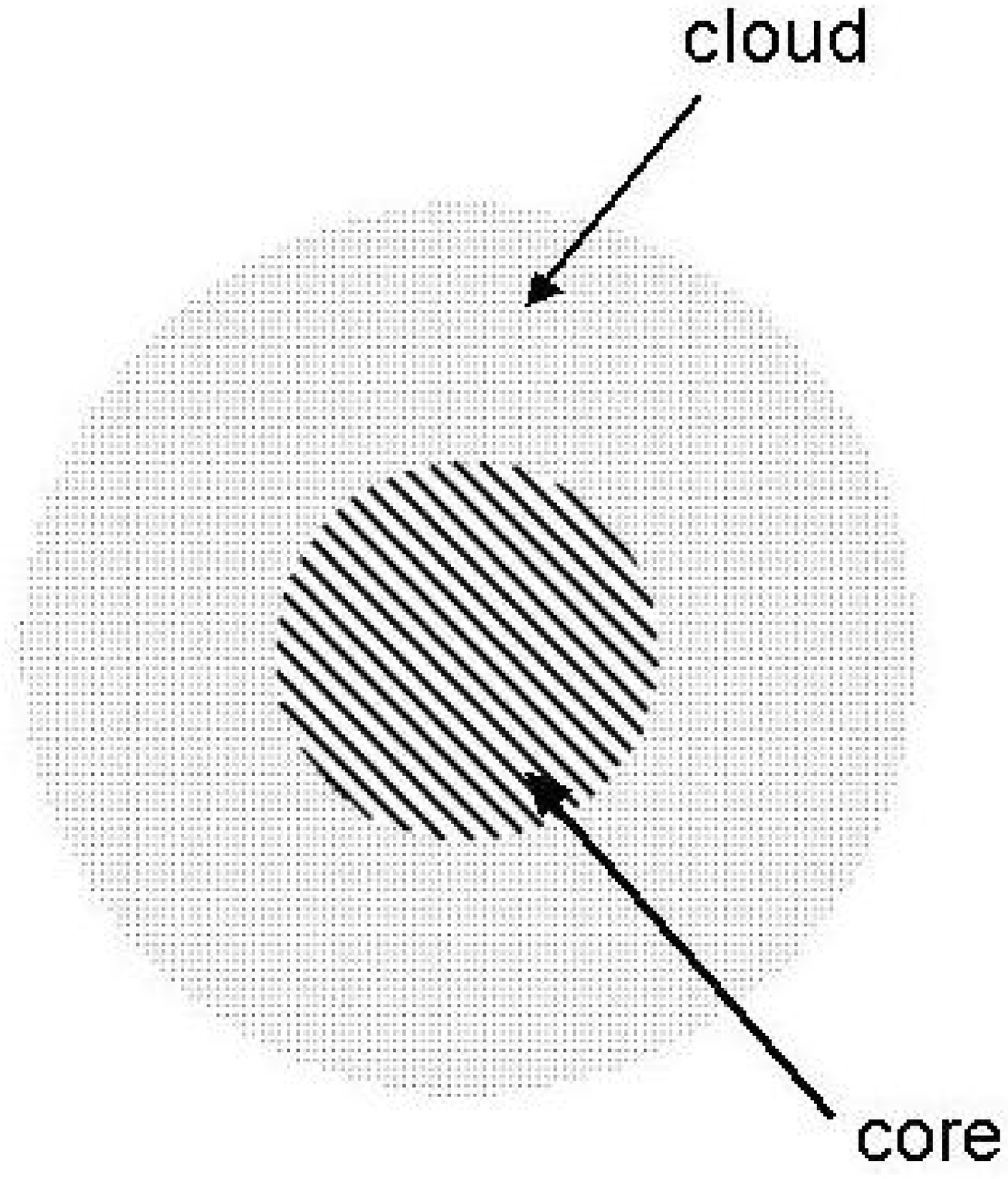}
  \hspace{1 cm}
  \includegraphics[height=.24\textheight]{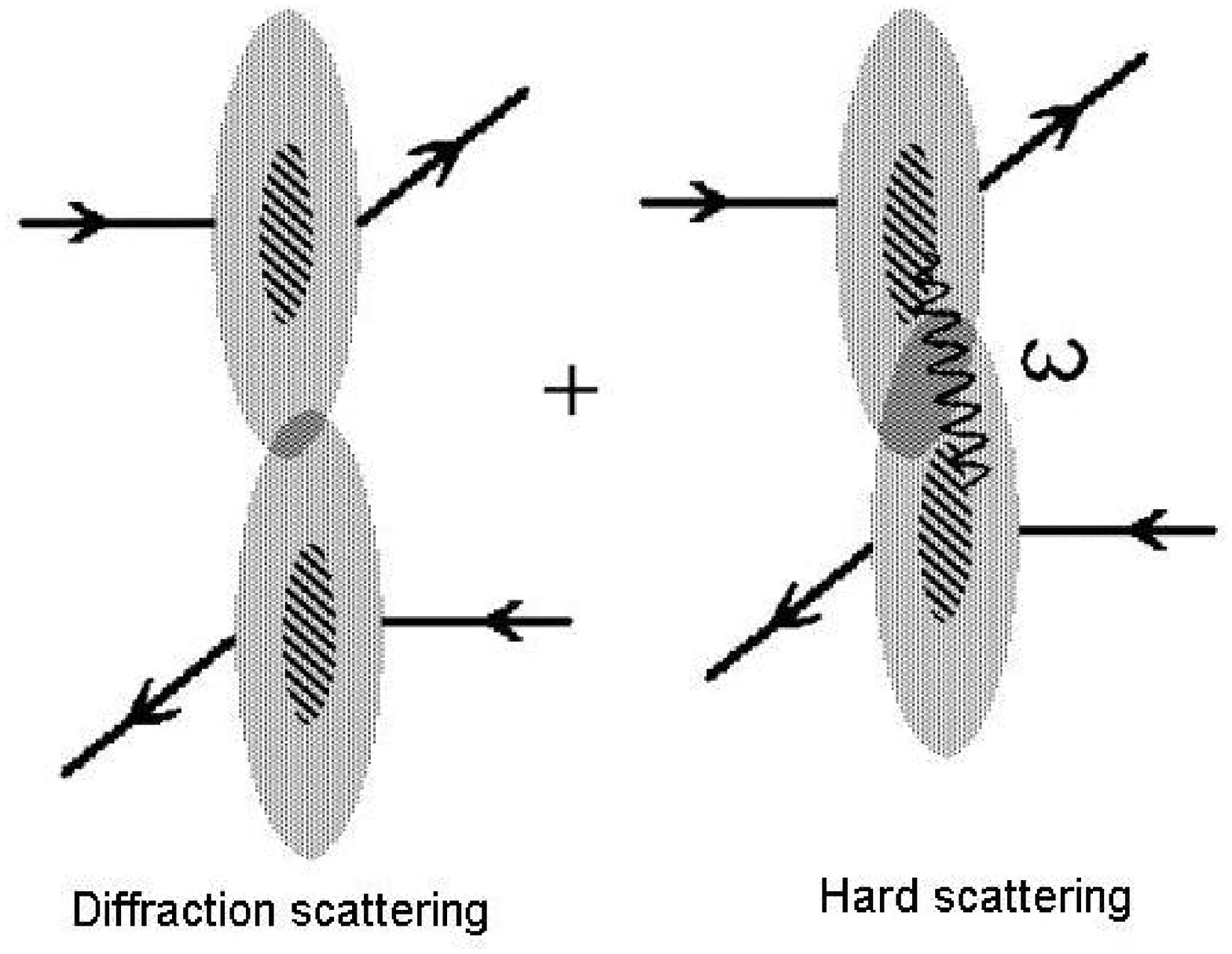}
  \caption{\textbf{\hspace{2in} FIGURE 2.}}
\end{figure}

\setcounter{figure}{2}

\begin{figure}[pt]
  \includegraphics[height=1.7in]{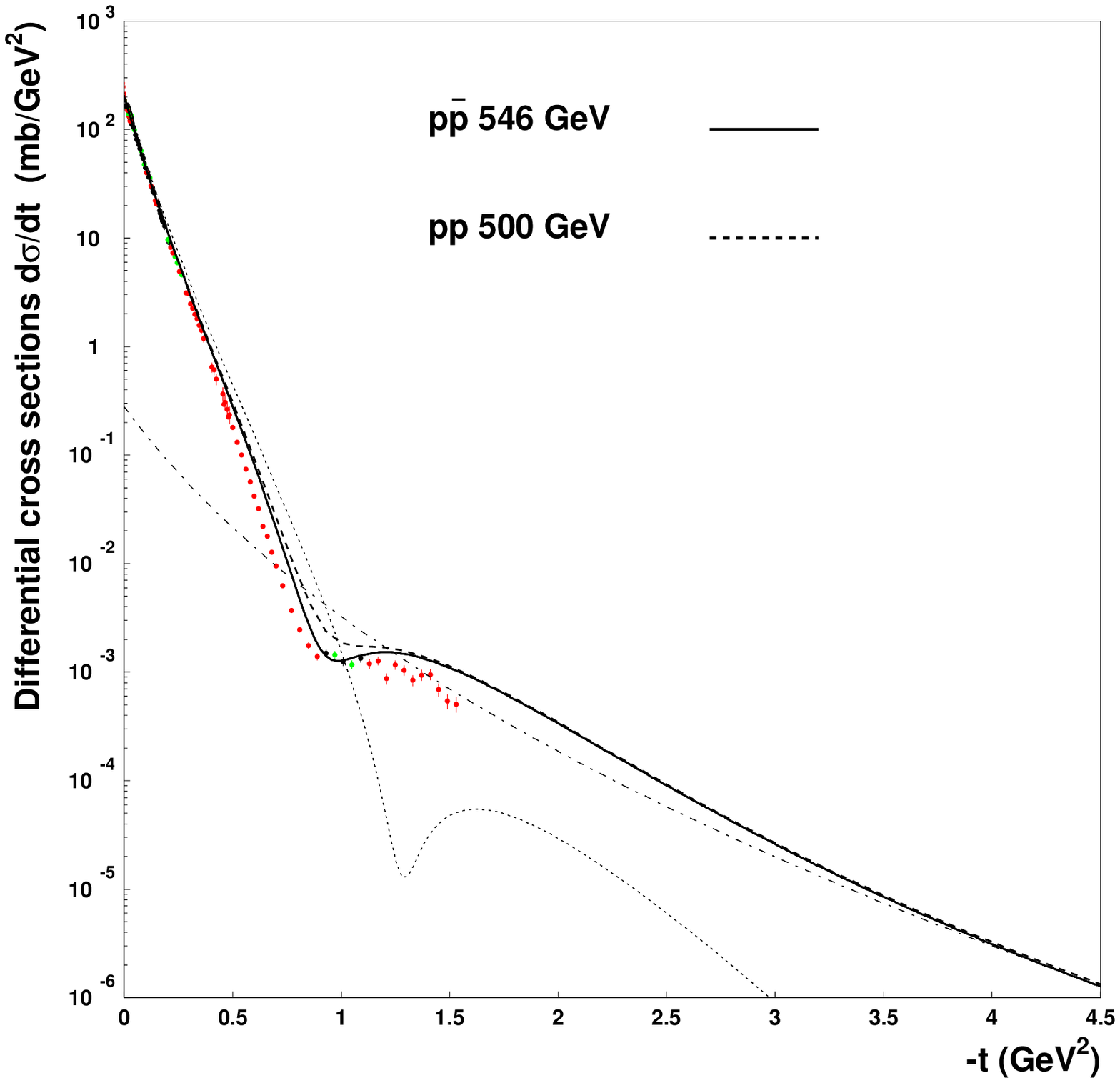}
  \hspace{.2 cm}
  \includegraphics[height=1.7in]{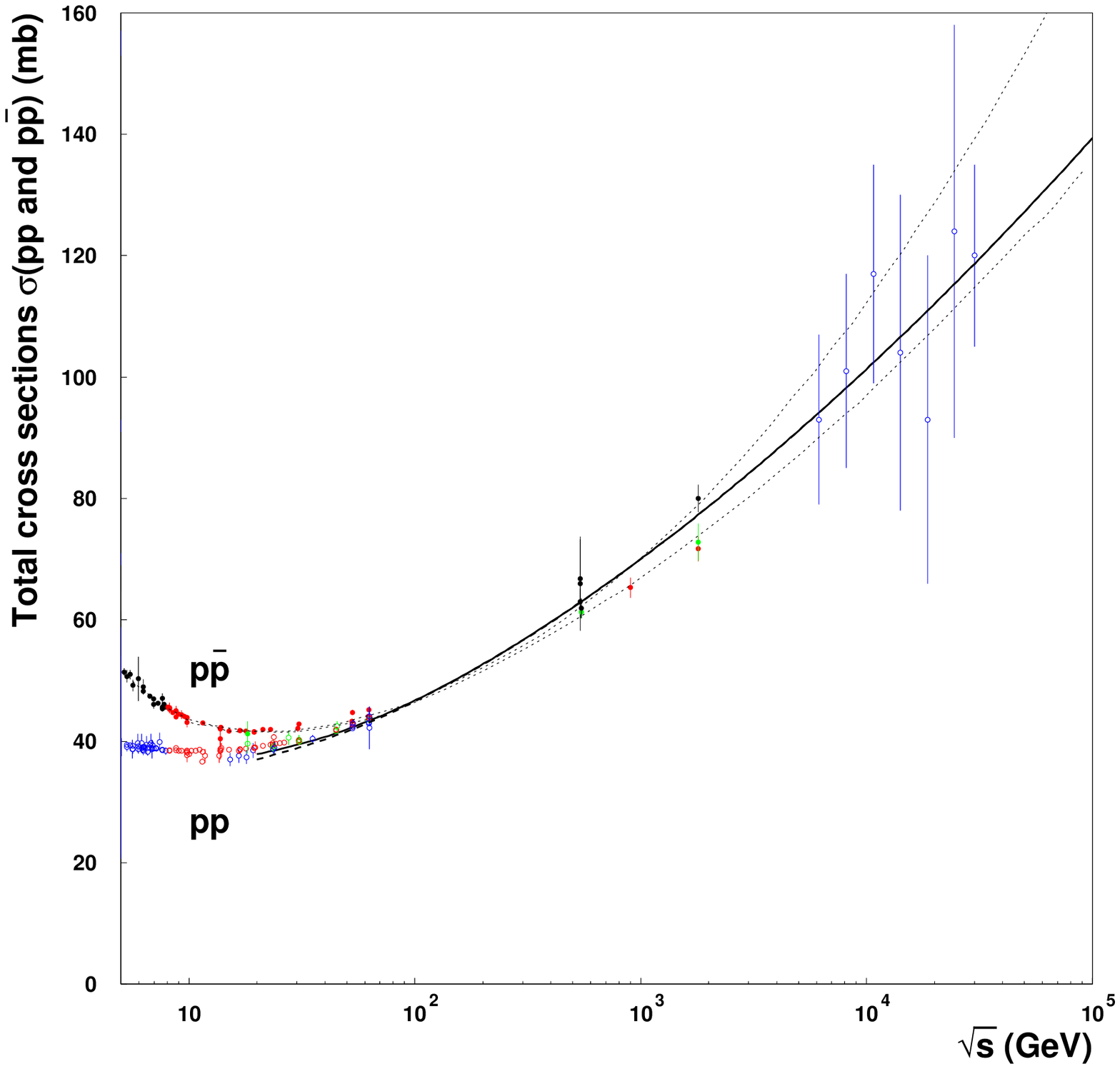}
  \hspace{.2 cm}
  \includegraphics[height=1.7in]{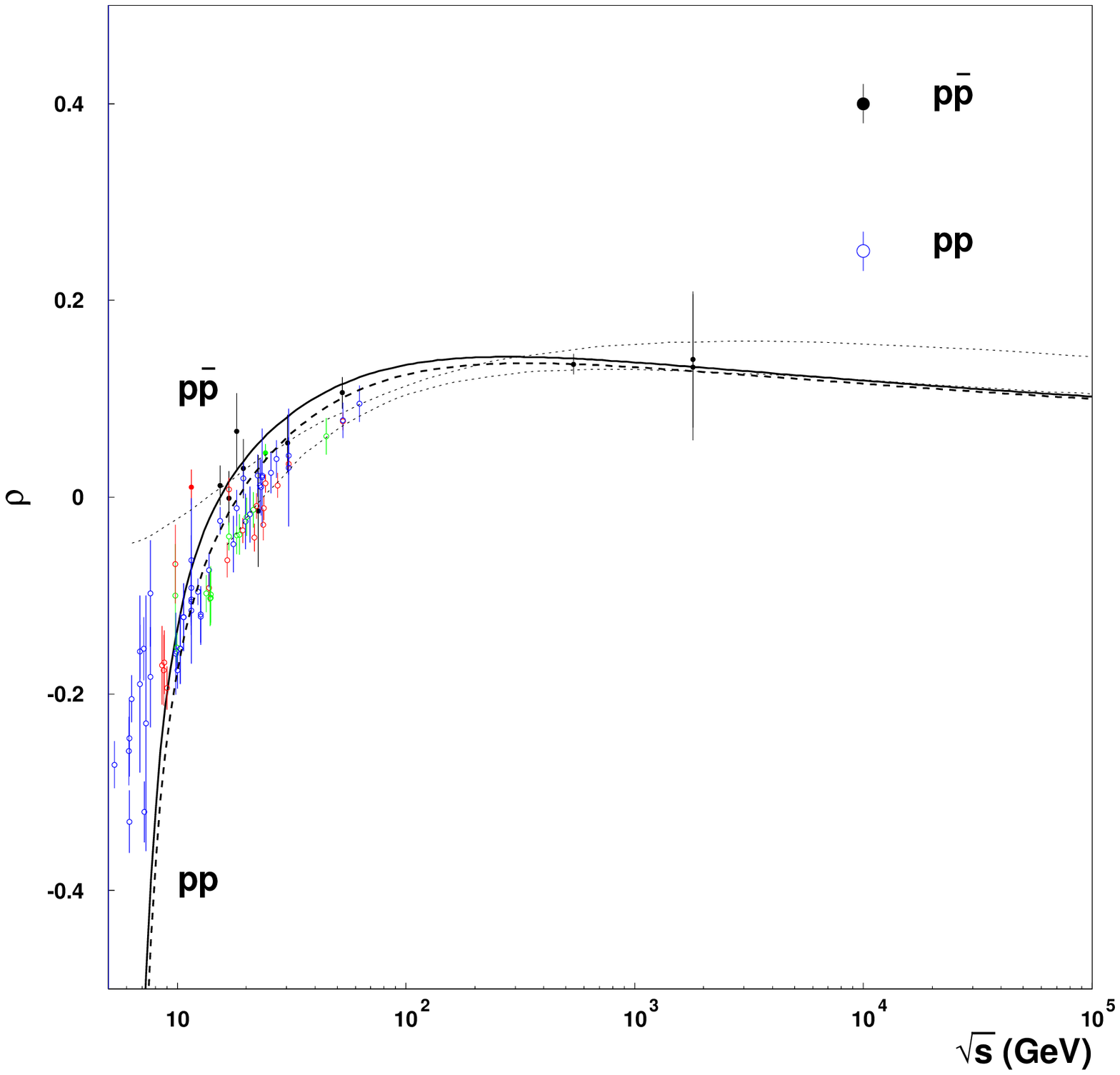}
  \caption{\textbf{\hspace{1in} FIGURE 4.\hspace{1in}FIGURE 5.}}
\end{figure}

\setcounter{figure}{5}

\begin{figure}[ht]
  \includegraphics[height=1.7in]{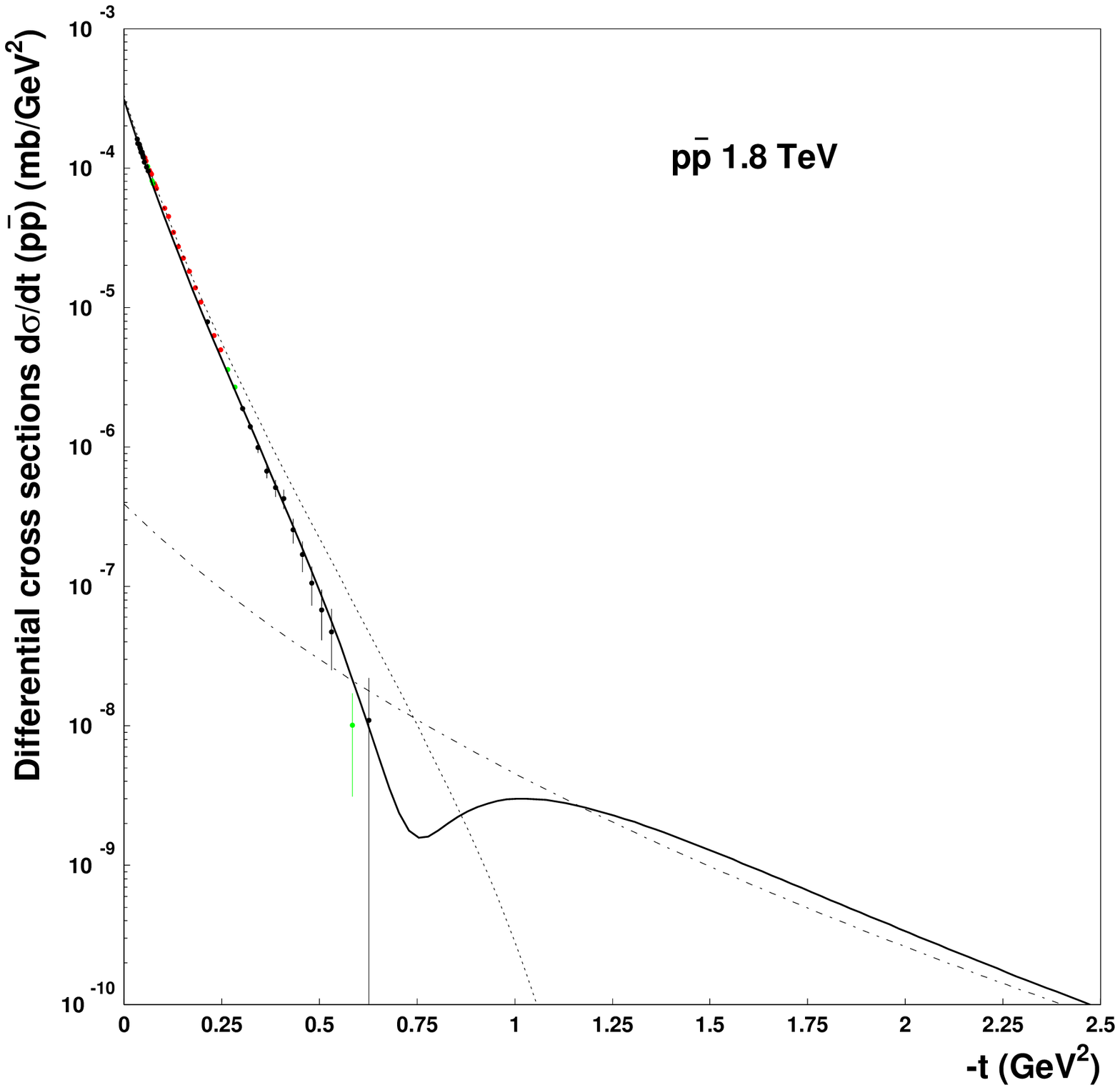}
  \hspace{.2 cm}
  \includegraphics[height=1.7in]{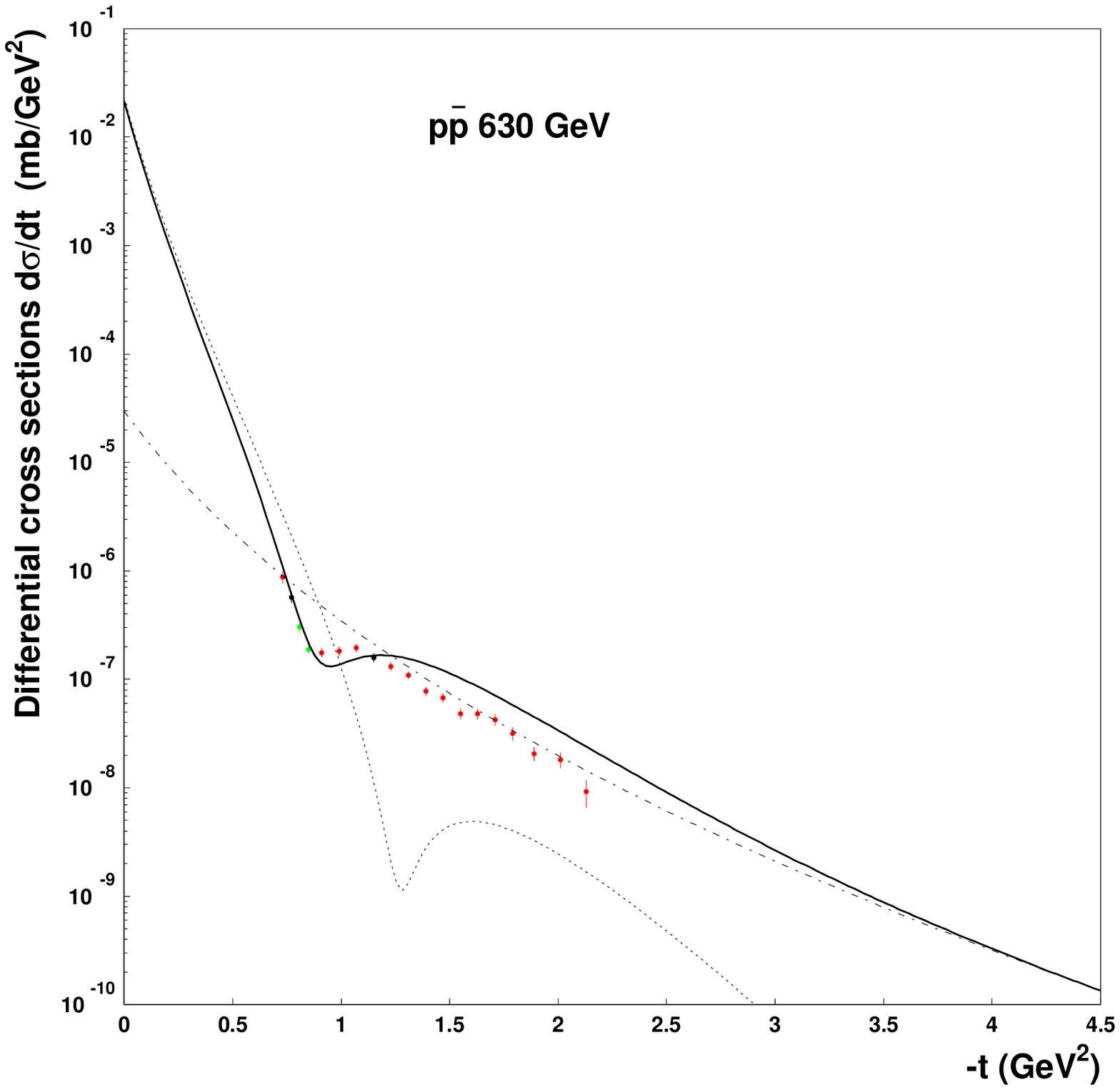}
  \hspace{.2 cm}
  \includegraphics[height=1.7in]{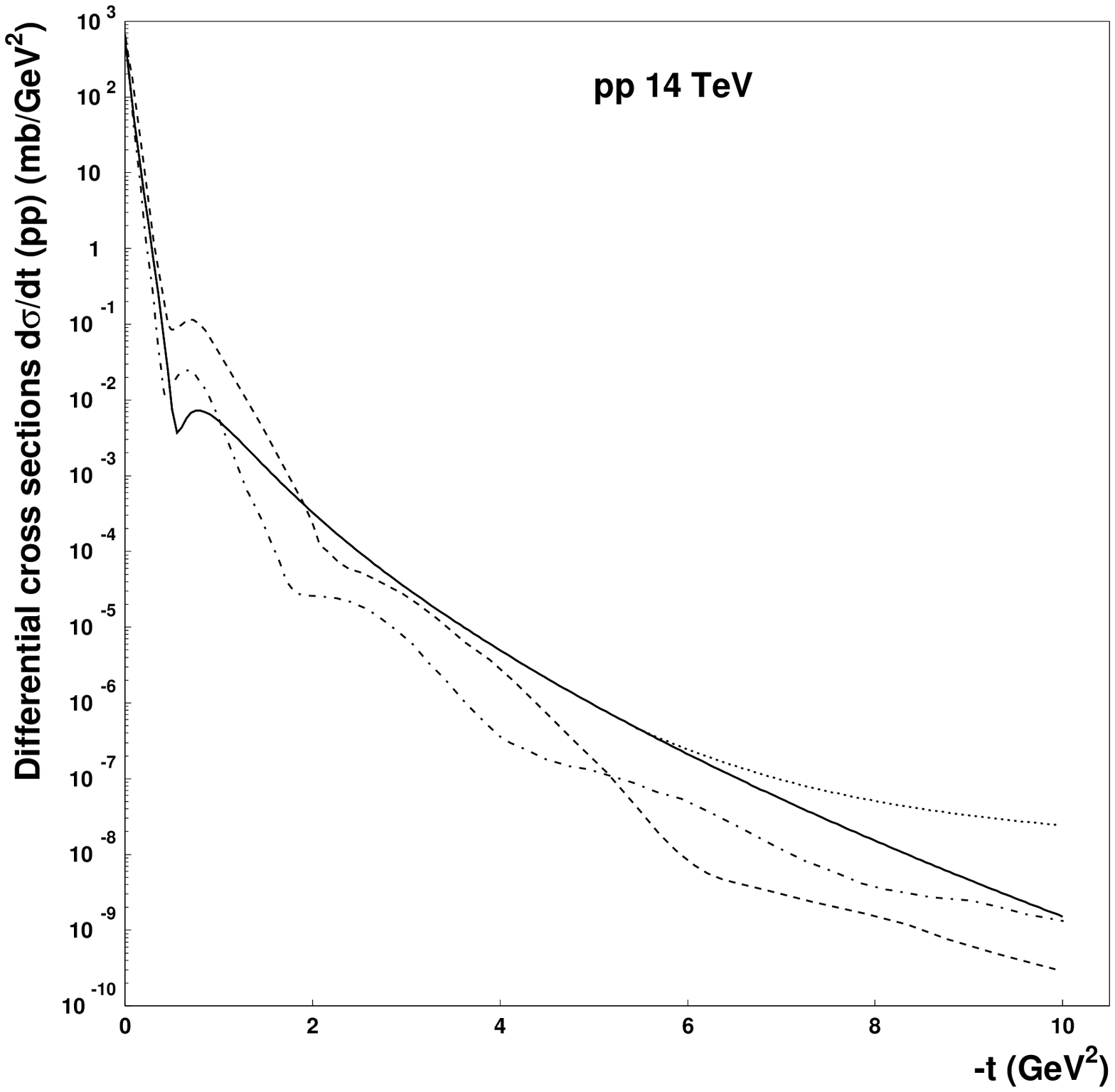}
  \caption{\textbf{\hspace{1in} FIGURE 7.\hspace{1in}FIGURE 8.}}
\end{figure}

\setcounter{figure}{8}

\begin{figure}[ht]
  \includegraphics[height=.17\textheight]{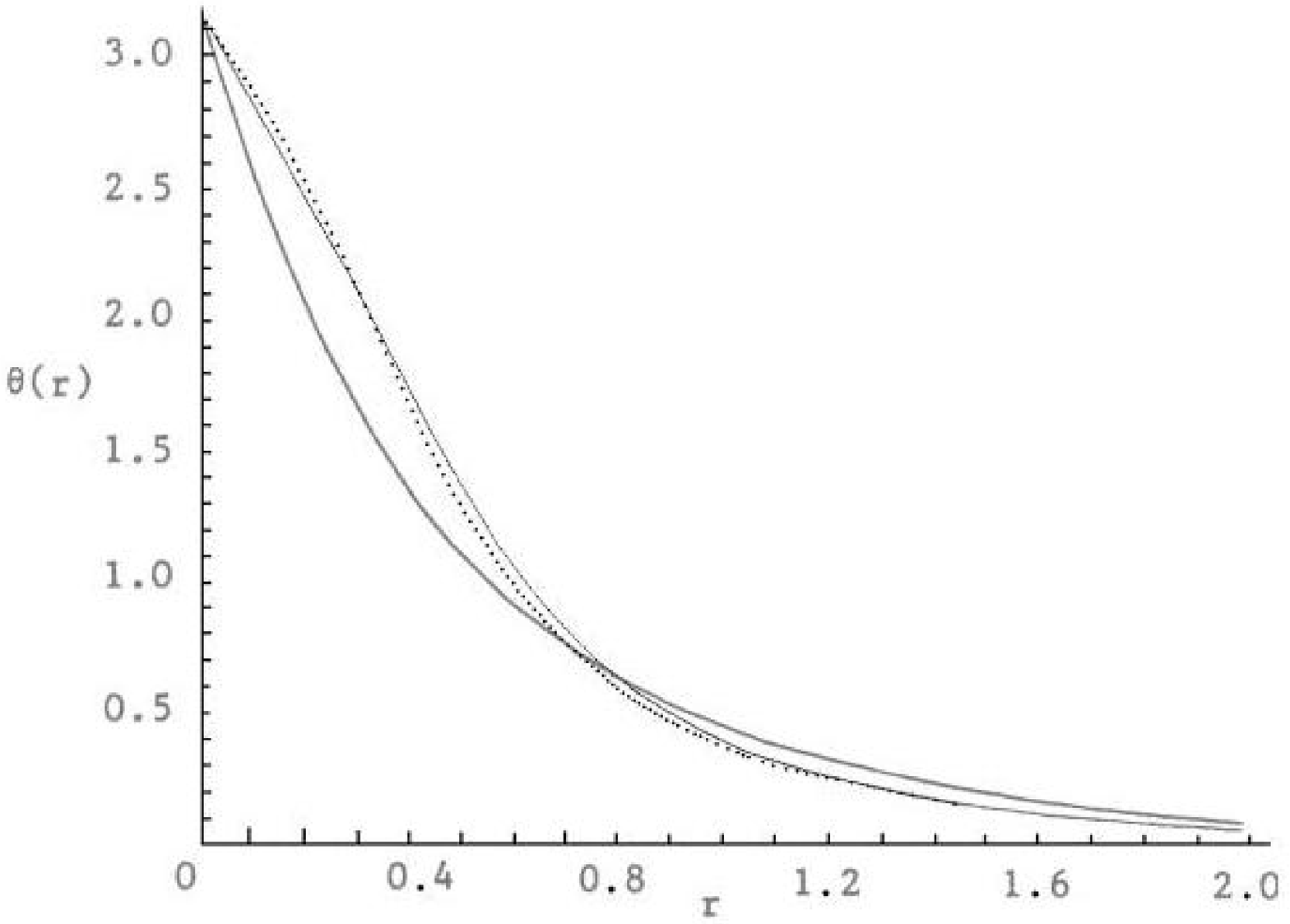}
  \hspace{2.5 cm}
  \includegraphics[height=.17\textheight]{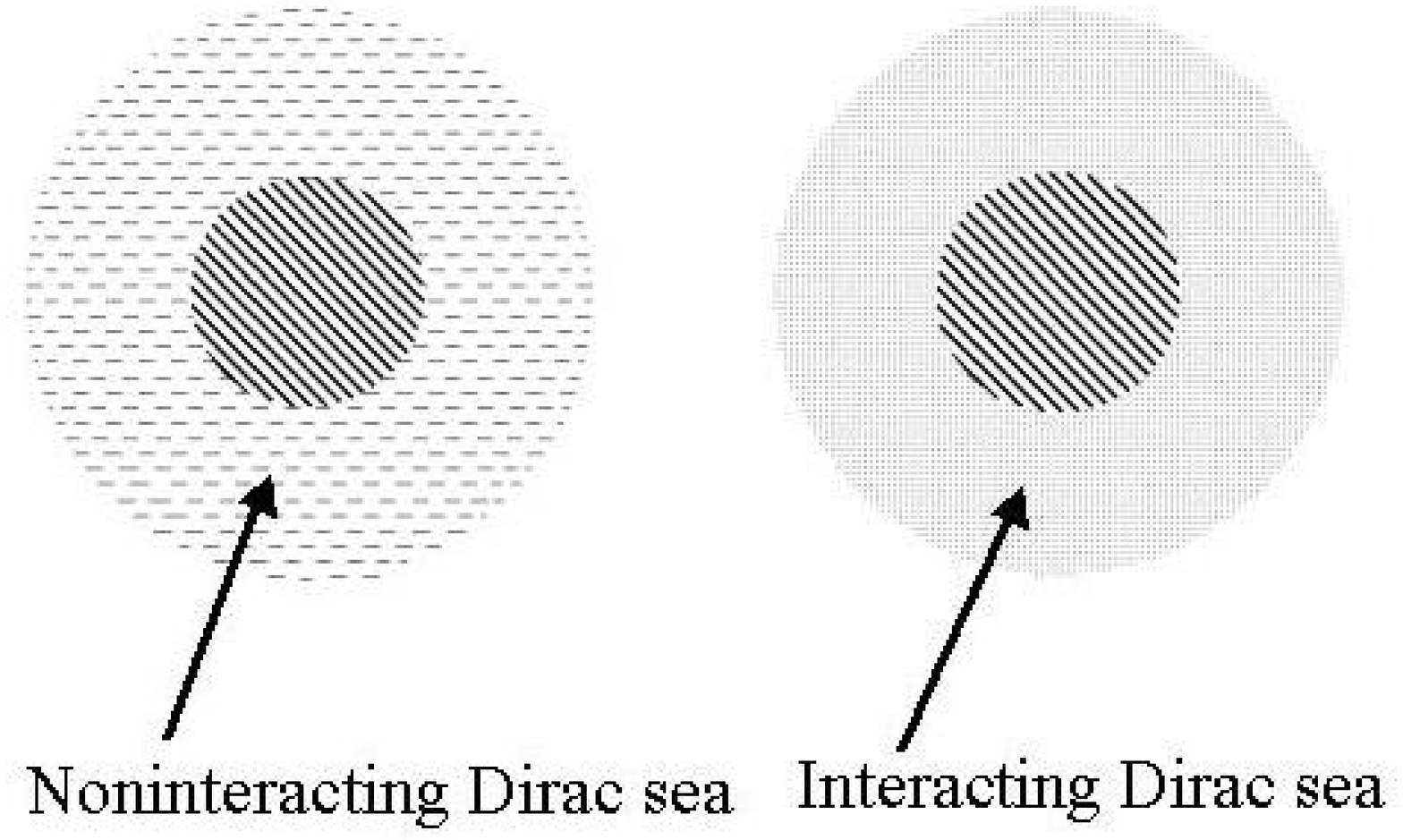}
  \caption{\textbf{\hspace{2in} FIGURE 10.}}
\end{figure}

\setcounter{figure}{10}

\begin{figure}[b]
  \includegraphics[height=1.9in,width=2.5in]{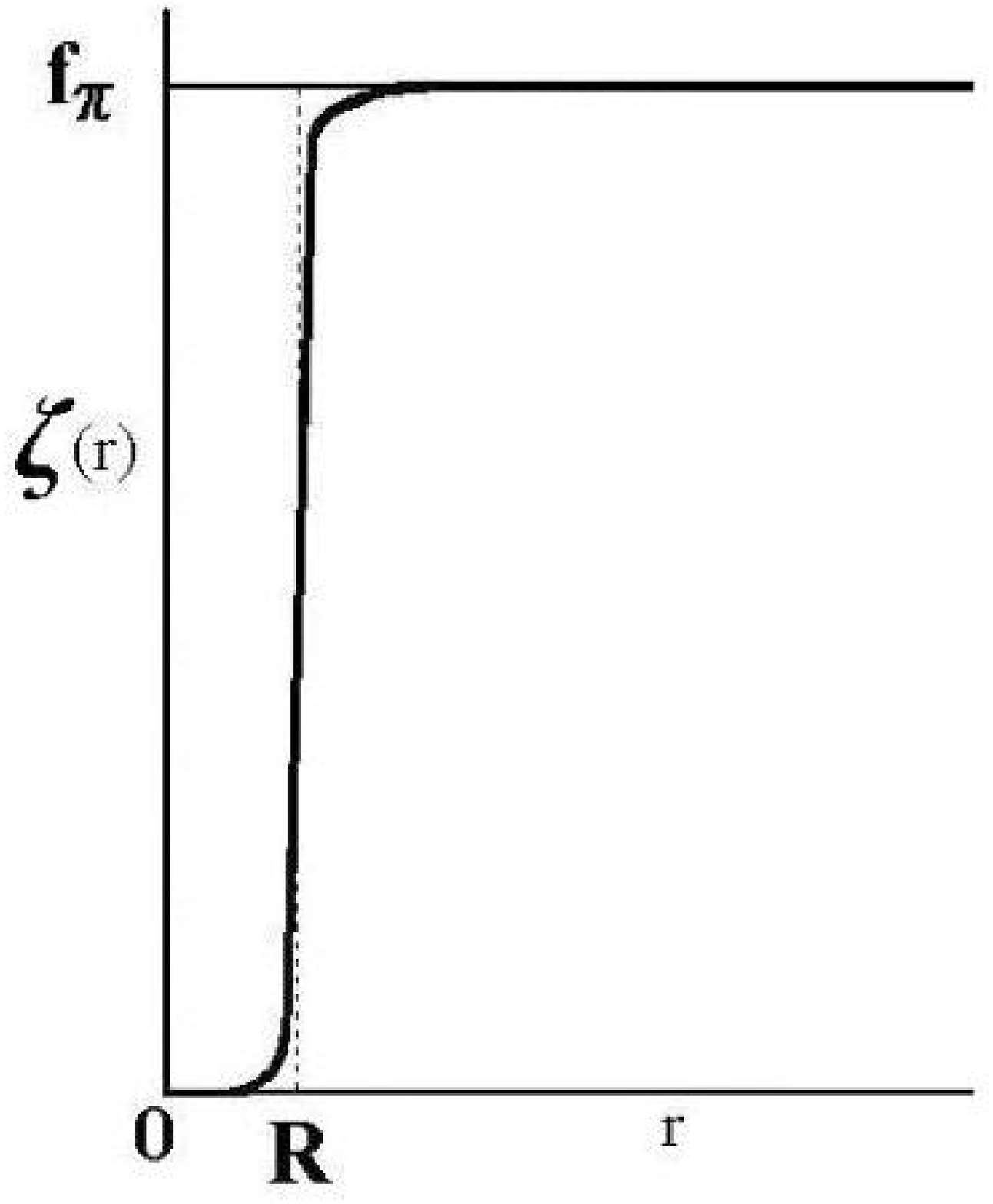}
  \hspace{1.3cm}
  \includegraphics[height=.22\textheight]{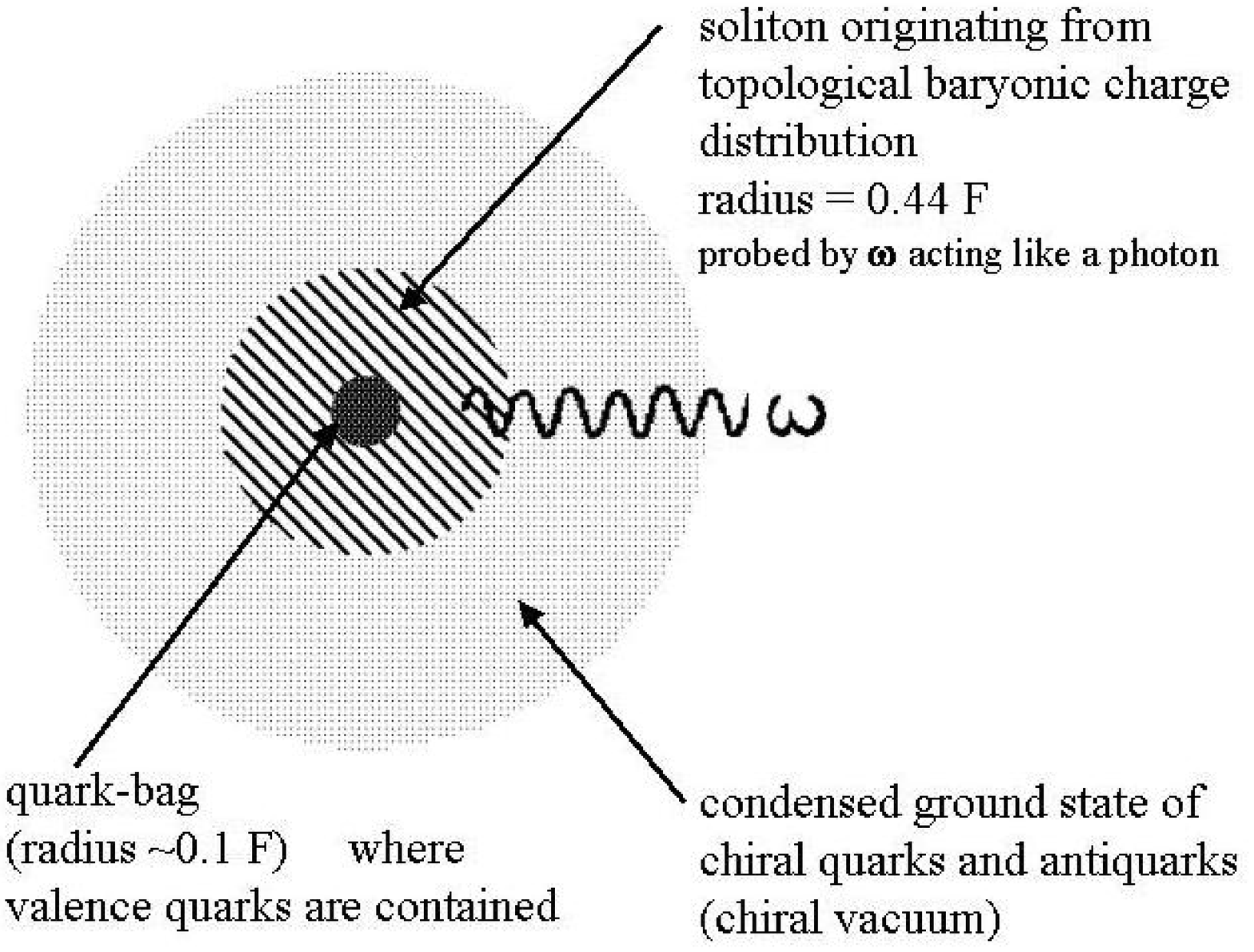}
  \caption{\textbf{\hspace{2in} FIGURE 12.}}

\end{figure}





\IfFileExists{\jobname.bbl}{}
 {\typeout{}
  \typeout{******************************************}
  \typeout{** Please run "bibtex \jobname" to optain}
  \typeout{** the bibliography and then re-run LaTeX}
  \typeout{** twice to fix the references!}
  \typeout{******************************************}
  \typeout{}
 }

\end{document}